\documentclass[a4paper,11pt]{article}
\usepackage{pos}

\usepackage{bm}
\usepackage{graphicx}
\usepackage{ascmac}
\usepackage{cancel}
\usepackage{braket}

\title{Compositeness of near-threshold states with repulsive Coulomb interaction combined with short-range potential}
\ShortTitle{Compositeness of near-threshold states with Coulomb plus short-range interactions}

\author*[a]{Tomona Kinugawa}
\author[b]{Tetsuo Hyodo}

\affiliation[a]{Nishina Center for Accelerator-Based Science, RIKEN, \\
  Wako 351-0198, Japan}

\affiliation[b]{Department of Physics, Tokyo Metropolitan University,\\
Hachioji 192-0397, Japan}

\emailAdd{tomona.kinugawa@riken.jp}
\emailAdd{hyodo@tmu.ac.jp}

\abstract{We investigate the internal structure of near-threshold states in a system with a repulsive Coulomb interaction combined with a short-range potential, using the compositeness. We construct a model in which the eigenmomentum is expressed in terms of three observables: the Coulomb scattering length, the Coulomb effective range, and the Bohr radius. In the presence of the Coulomb interaction, a bound state directly goes into a resonance as parameters are varied, bypassing a virtual state, in contrast to the case with only the short-range interaction. We show that the compositeness of near-threshold states can be expressed solely in terms of these observables. When the magnitude of the Coulomb effective range is much smaller than that of the Bohr radius, both shallow bound states and near-threshold resonances exhibit common structures with large compositeness, reflecting the remnant of the low-energy universality.}

\FullConference{The 21st International Conference on Hadron Spectroscopy and Structure (HADRON2025)\\
 27 - 31 March, 2025\\
Osaka University, Japan\\}

\begin{document}
\maketitle



\section{Introduction}

Understanding the internal structure of exotic hadrons is a central goal in hadron physics. There have been several discoveries of exotic hadrons in the heavy sector, which have attracted significant attention recently~\cite{ParticleDataGroup:2024cfk}. The internal structure of exotic hadrons is considered to be different from that of ordinary mesons, which consist of a quark-antiquark pair, or baryons composed of three quarks. With four or more quarks, various structures are possible for exotic hadrons. For example, hadronic molecules, which are weakly bound states of two hadrons, and multiquark states, which are compact configurations of several quarks, have been proposed. A lot of studies have been carried out to understand the internal structure of exotic hadrons~\cite{Hosaka:2016pey,Guo:2017jvc,Brambilla:2019esw,Hosaka:2025gcl}. 

In practice, the internal structure of exotic hadrons is expressed as the superposition of possible components. The weight of the hadronic molecule component is called the compositeness. More precisely, the compositeness $X$ is defined as the probability of finding the free scattering state $\ket{\bm{p}}$ in the bound state $\ket{B}$~\cite{Hyodo:2013nka,vanKolck:2022lqz,Kinugawa:2024crb}:
\begin{align}
X &= \int \frac{d\bm{p}}{(2\pi)^{3}}\ |\braket{\bm{p}|B}|^{2}.
\end{align}
The remaining probability $Z = 1 - X$ is called the elementarity, the probability of finding the non-molecular components. By evaluating the compositneess $X$, we can quantitatively analyze the internal structure of exotic hadrons from the perspective of hadronic molecules. 

For the study of the internal structure of exotic hadrons, we focus on the fact that the majority of exotic hadrons are observed near the threshold of two-body scatterings. If the scatterings occur with only the $s$-wave short-range interaction, it is shown that the low-energy universality emerges in the near-threshold energy region~\cite{Braaten:2004rn,Naidon:2016dpf}. As a consequence of the universality, shallow bound states usually have the composite dominant structure~\cite{Kinugawa:2022fzn,Kinugawa:2023fbf}, while the near-threshold resonances tend to be non-composite dominant states~\cite{Hyodo:2013iga,Matuschek:2020gqe,Kinugawa:2024kwb}. In this way, the universal nature of near-threshold $s$-wave states with short-range interaction is established.

However, in the presence of a long-range interaction, such as the Coulomb force, the situation can be qualitatively different. It is known that the low-energy scattering amplitude of the Coulomb plus short-range interactions differs from that of the short-range interaction without the Coulomb interaction~\cite{Bethe:1949yr,Domcke:1983zz,Kong:1998sx,Kong:1999sf,Ando:2007fh,Higa:2008dn,Mochizuki:2024dbf}. Although the Coulomb interaction is much weaker than the short-range interactions in the typical hadron systems, it becomes important in the near-threshold energy region. For example, the lattice QCD suggests that the $\Omega_{ccc}^{++}$-$\Omega_{ccc}^{++}$ system forms a shallow bound state with only the strong interaction, but it becomes an unbound state when the Coulomb repulsion is switched on~\cite{Lyu:2021qsh}. Similarly, in the nuclear systems, ${}^{8}$Be nucleus is a near-threshold resonance in two $\alpha$ scatterings, while only the short-range interaction forms an $\alpha\alpha$ bound state without Coulomb repulsion~\cite{Braaten:2004rn,Higa:2008dn}. In this way, the Coulomb interaction plays a significant role in determining the nature of near-threshold states.

As mentioned above, in systems with only short-range interactions, the structure of near-threshold states is completely different depending on their position, whether it is above (a resonance) or below (a bound state) the threshold. Therefore, it is not clear whether the internal structure remains unchanged when a bound state turns into a resonance due to the Coulomb repulsion. Based on this background, we investigate the internal structure of near-threshold states in systems with the Coulomb plus short-range interactions. In this work, we focus only on the repulsive Coulomb plus short-range interaction. 


\section{Model for Coulomb plus short-range interaction}
\label{sec:model}

To consider a system with the Coulomb plus short-range interactions, we introduce a model in which two particles with charge $Z_{1}e$ and $Z_{2}e$ scatter and couple to a discrete state~\cite{Domcke:1983zz}. In this model, the self-energy of the discrete state $\Sigma(E)$ with repulsive Coulomb interaction is obtained as~\cite{Domcke:1983zz},
\begin{align}
\Sigma(E) = -\frac{A}{2\pi}\left[C + \frac{1}{2}ia_{B}k + \log(-ia_Bk) + \psi\left(1+\frac{i}{a_Bk}\right) \right],
\end{align}
where $k = \sqrt{2\mu E}$ is the momentum with the reduced mass $\mu$. The model parameter $A$ corresponds to the strength of the coupling, and $C$ determines the constant part of the self-energy. The digamma function is defined as $\psi(x) = d\log\Gamma(x)/dx $. We define the Bohr radius $a_{B}$ both for attractive and repulsive interactions as
\begin{align}
a_{B} &= \frac{1}{\mu \alpha |Z_1 Z_2|},
\end{align}
where $\alpha\approx 1/137$ is the fine-structure constant. $a_{B}$ characterizes the strength of the Coulomb interaction. For a smaller (larger) $a_{B}$, the Coulomb interaction is stronger (weaker). 

In general, the eigenmomentum is determined by the pole of the scattering amplitude. In the near-threshold region, that pole condition $E - \nu_{0} - \Sigma(E)=0$ with the bare mass $\nu_{0}$ can be rewritten in terms of the observables~\cite{Bethe:1949yr}
\begin{align}
-\frac{1}{a_{s}} + \frac{r_{e}}{2}k^{2} - ik - \frac{2}{a_B}\left [\log(-ia_Bk) + \psi\left(1 + \frac{i}{a_Bk}\right)\right] &= 0.
\label{eq:pole}
\end{align}
$a_{s}$ and $r_{e}$ are the Coulomb scattering length and Coulomb effective range, which characterize the property of the short-range interaction. By solving this equation for complex $k$, the eigenmomentum $k_{h}$ can be obtained. When the pole is on the positive imaginary axis (${\rm Re}\ k_{h}=0$, ${\rm Im}\ k_{h}>0$), it represents a bound state. On the other hand, a pole corresponding to a resonance is in the fourth quadrant (${\rm Re}\ k_{h}>0$, ${\rm Im}\ k_{h}<0$)~\cite{Taylor}. The pole condition~\eqref{eq:pole} is also derived within the framework of the effective field theory~\cite{Kong:1998sx,Kong:1999sf,Ando:2007fh,Higa:2008dn}. When the momentum is sufficiently small such that the $k^{2}$ term can be neglected, it is shown that the condition only depends on $a_{s}$ and $a_{B}$~\cite{Mochizuki:2024dbf}. As shown in Eq.~\eqref{eq:pole}, the logarithmic and digamma terms appear in the presence of the Coulomb interaction, in contrast to the short-range case. Due to the $\log(-ia_Bk)$ term, the function in the left-hand side of Eq.~\eqref{eq:pole} exhibits a branch cut along the negative imaginary axis in the complex $k$ plane. This feature leads to distinct pole trajectory behavior compared to short-range systems. 

The compositeness $X$ of eigenstates is calculated from the energy derivative of the self-energy $\Sigma(E)$~\cite{Hyodo:2014bda,Kinugawa:2024crb}:
\begin{align}
X &= \left.\frac{-\Sigma'(E)}{1 - \Sigma'(E)}\right|_{E = E_{h}},
\label{eq:X}
\end{align}
where we denote $d\Sigma(E)/dE = \Sigma'$, and $E_{h}=k_{h}^{2}/(2\mu)$ stands for the eigenenergy. From Eq.~\eqref{eq:X}, the compositeness $X$ can be expressed in terms of observables:
\begin{align}
X &= \left[1 - \frac{r_{e}}{R}\right]^{-1},
\label{eq:wbr}\quad
R = -\frac{1}{k_{h}}\left[2i\left(\frac{1}{a_{B}k_{h}}\right)^{2}\psi_{1}\left(i\frac{1}{a_{B}k_{h}}\right)+ i - 2\left(\frac{1}{a_{B}k_{h}}\right) \right],
\end{align}
with the trigamma function $\psi_{1}(x)=d^{2}\log \Gamma(x)/dx^{2}$. The contribution of $a_{s}$ is contained in this relation through the pole condition Eq.~\eqref{eq:pole}.

\section{Properties of near-threshold states}
\label{sec:property}

\begin{figure}[tbp]
\centering
\includegraphics[width=0.45\textwidth]{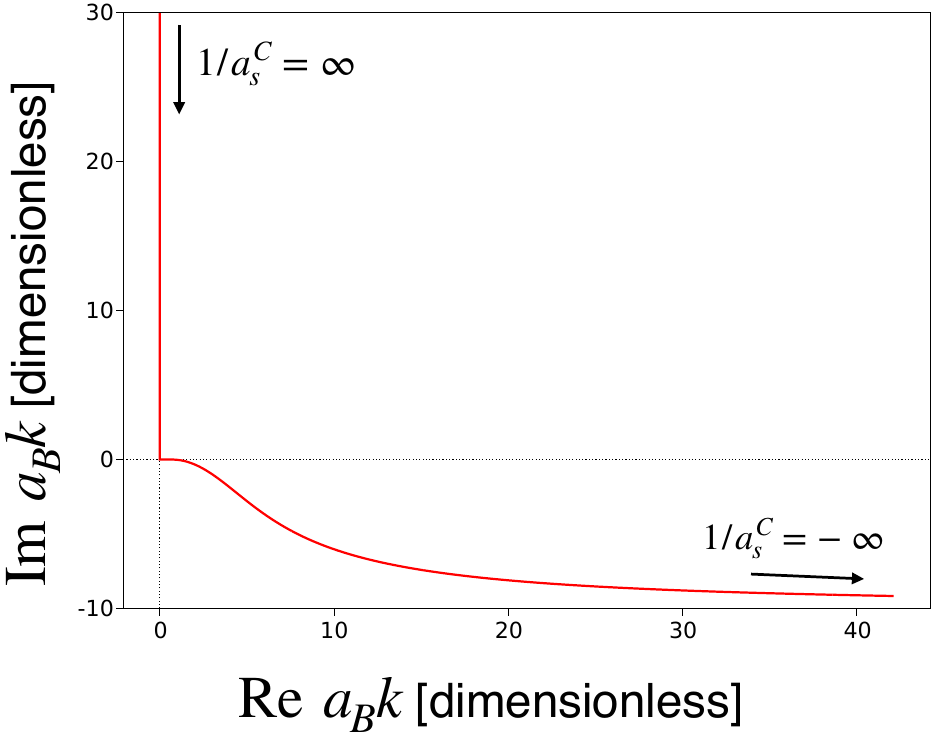}
\includegraphics[width=0.45\textwidth]{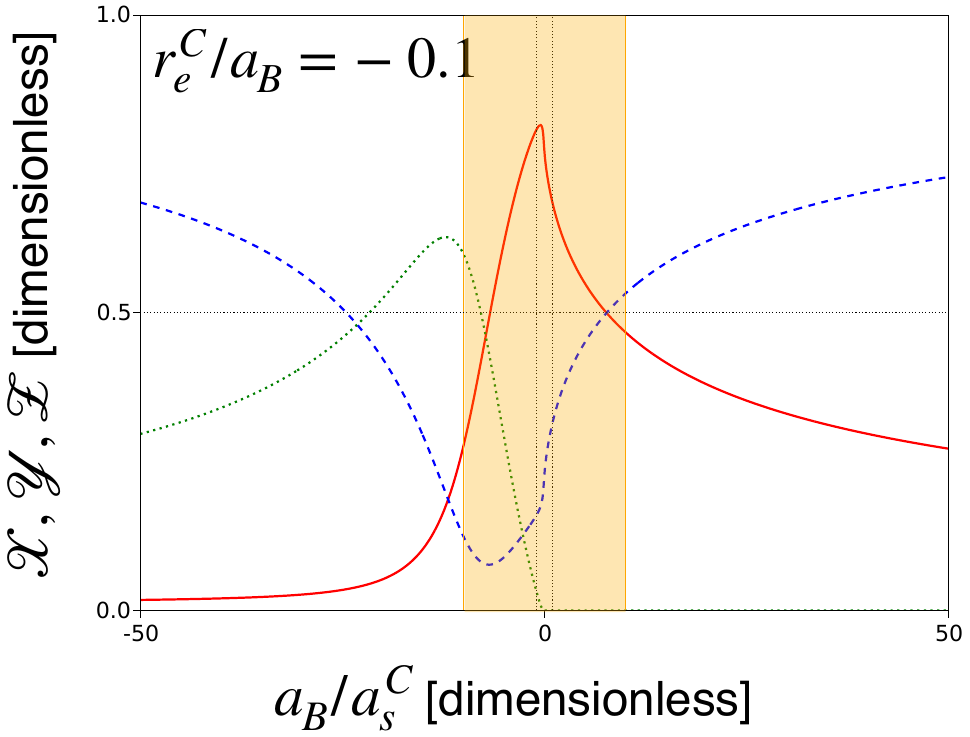}
\caption{Left panel: The pole trajectory by varying $a_{B}/a_{s}$ with negative $r_{e}/a_{B} = -0.1$. Right panel: The probabilities $\mathcal{X}$ (solif line), $\mathcal{Y}$ (dotted line), and $\mathcal{Z}$ (dashed line) as functions of $a_{B}/a_{s}$ with $r_{e}/a_{B} = -0.1$. 
The shaded area shows the $|a_{B}/a_{s}| < |a_{B}/r_{e}|$ region, and the virtical dotted lines correspond to $|a_{B}/a_{s}| = 1$.}
\label{fig:pole-XYZ}
\end{figure}

Here we discuss the nature of near-threshold states with repulsive Coulomb plus short-range interactions numerically. In the left panel of Fig.~\ref{fig:pole-XYZ}, we show the pole trajectory from the bound state to the resonance by varying the scattering length $a_{s}$ with keeping the effective range fixed at $r_{e}/a_{B} = -0.1$. For simplicity, only the right half of the momentum plane is shown, as the left half is symmetric due to the Schwarz reflection principle~\cite{Taylor}.

When $a_{B}/a_{s}$ is positive and large, a deep bound state is formed, whose pole is located on the imaginary $k$ axis. By decreasing $a_{B}/a_{s}$, the bound state becomes shallower. In the $a_{B}/a_{s} \to 0$ limit, the binding energy goes to zero and the pole reaches the threshold $k = 0$. When $a_{B}/a_{s}$ becomes negative, the pole moves to the fourth quadrant of the $k$ plane. In other words, the bound state directly turns into a resonance without passing through a virtual state. This behavior is completely different from that in a system with only short-range interactions. This result indicates that the near-threshold states do not necessarily have to be composite dominant, in contrast to the short-range systems. In fact, it is known that the low-energy universality is not realized in this case because the Bohr radius is non-negligible~\cite{Domcke:1983zz,Mochizuki:2024dbf}.

To investigate the internal structure, we evaluate the compositeness $X$. However, the compositeness of resonance with negative $a_{B}/a_{s}$ is defined as complex, which cannot be regarded as a probability. For probabilistic interpretation, we adopt the scheme proposed in Ref.~\cite{Kinugawa:2024kwb}. In this scheme, the fraction of the composite (elementary) component of resonance is characterized by the probabilistic quantity $\mathcal{X}$ ($\mathcal{Z}$), which is calculated from the complex compositeness $X$. Furthermore, $\mathcal{Y}$ is defined to quantify the ambiguity in identifying the nature of the resonance. For a bound state with $0\leq X\leq 1$, we have $\mathcal{X}=X$, $\mathcal{Y}=0$, and $\mathcal{Z}=1-X$. In the right panel of Fig.~\ref{fig:pole-XYZ}, we show $\mathcal{X,Y,Z}$ of the pole as functions of $a_{B}/a_{s}$.

When the $|a_{B}/a_{s}|$ is large, namely, the pole exists in the far-threshold region, the compositeness $\mathcal{X}$ is small, reflecting that the eigenstate originates from a pure elementary state with $X = 0$ in this model. On the other hand, in the near-threshold region with small $|a_{B}/a_{s}|$, especially in the $|a_{B}/a_{s}| < |a_{B}/r_{e}|$ region (indicated by the shaded region), the compositeness approaches unity. This can be regarded as the remnant of the low-energy universality, which enlarges the compositeness $\mathcal{X}$ near the threshold before the Coulomb interaction becomes sizable. However, when the pole goes further close to threshold within $|a_{B}/a_{s}| < |a_{B}/a_{B}|$ region (indicated by the virtical dotted line), the Coulomb interaction strongly affects the nature of the near-threshold states, and the compositeness remains $\mathcal{X} < 1$. Because of the remnant of the low-energy universality, the structure of near-threshold resonances is similar to that of the shallow bound states with large $\mathcal{X}$. 

\section{Summary}

In this study, we investigate the nature of near-threshold states in a system with repulsive Coulomb plus short-range interactions using the compositeness. It is shown that the compositeness of near-threshold states can be expressed only by the observables. We find that as the scattering length varies, a shallow bound state directly evolves into a resonance, which is qualitatively different from the short-range systems. While near-threshold states are not necessarily composite dominant in the presence of the Coulomb interaction, we show that when the effective range is smaller than the Bohr radius, the near-threshold states become composite dominant due to the remnant of low-energy universality.



\end{document}